# Quality of service monitoring:

# Performance metrics across proprietary content domains

Shawn O'Donnell[1]
Hugh Carter Donahue[2]
Josephine Ferrigno-Stack[3]

We propose a quality of service (QoS) monitoring program for broadband access to measure the impact of proprietary network spaces. Our paper surveys other QoS policy initiatives, including those in the airline, and wireless and wireline telephone industries, to situate broadband in the context of other markets undergoing regulatory devolution. We illustrate how network architecture can create impediments to open communications, and how QoS monitoring can detect such effects. We present data from a field test of QoS-monitoring software now in development. We suggest QoS metrics to gauge whether information "walled gardens" represent a real threat for dividing the Internet into proprietary spaces.

To demonstrate our proposal, we are placing our software on the computers of a sample of broadband subscribers. The software periodically conducts a battery of tests that assess the quality of connections from the subscriber's computer to various content sites. Any systematic differences in connection quality between affiliated and non-affiliated content sites would warrant research into the behavioral implications of those differences. If, however, the data shows that there are no chronic biases in connection quality, then it would be fair to conclude that the walls on the garden are low enough not to be detrimental to public communications.

QoS monitoring is timely because the potential for the Internet to break into a loose network of proprietary content domains appears stronger than ever. Recent court rulings and policy statements suggest a growing trend towards relaxed scrutiny of mergers and the easing or elimination of content ownership rules. This policy environment could lead to a market with a small number of large, vertically integrated network operators, each pushing its proprietary content on subscribers.

---

[1] Fletcher School of Law and Diplomacy, Tufts University; MIT Program on Internet Telecoms convergence. `sro@itc.mit.edu`

[2] Philadelphia, Pennsylvania. `hdonahue@pobox.asc.upenn.edu`

[3] Annenberg School for Communications, University of Pennsylvania. `jferrigno@asc.upenn.edu`

The move towards proprietary space conflicts with the open philosophy on which the Internet was founded. That alone, however, is not a reason for regulators to intervene. Policy makers require empirical evidence that proprietary barriers require a public response. Unfortunately, traditional indicators of harm to consumers from industry mergers, like measures of programming diversity on cable systems, are insufficient when Internet connectivity becomes the norm for access networks.

Conventional measures of diversity make little sense if all providers offer access to the Internet. Any differences in information variety resulting from special, proprietary content would be swamped by the content of the Internet. What matters more than a binary measure of availability are the quality and equality of the connection to diverse content providers over access networks.

It is unlikely that an Internet access provider would completely block the content of its competitors. Any provider wishing to steer subscribers away from non-affiliated content would be more likely to do so by delivering that content at a slightly lower quality compared to affiliated content. The degradation of quality need not be blatant to be effective--a differential of a few milliseconds between affiliated and unaffiliated sites should suffice to condition users to abandon non-affiliated, "slower" content sources. A QoS monitoring system would alert policymakers to the development of such scenarios.

## 1.  Proprietary information spaces on the Internet: What's the problem?

Shakeouts in the Internet access industry and mergers in the media industry are reducing the number of options available to American consumers.  Moreover, vertically integrated conglomerates such as AOL Time Warner raise the potential for "walled gardens" or proprietary information spaces.  In such closed spaces, the owner of the communications infrastructure tilts the competition among content providers in the direction of providers owned by or affiliated with the infrastructure owner.  Through technical or other means, it is possible for the infrastructure provider to make it difficult or impossible to access information from providers outside the walled garden.

Why should anyone care if there are proprietary information spaces on the Internet? For most consumers, the problem with proprietary information spaces is that they could effectively restrict choices among information sources. The presence of competition among vertically integrated carriers and information sources can mislead policy analysts into a false sense of security. First, it is important to distinguish here between the competition in markets as a whole and choice for individual consumers. While there might be several cable operators in a major metropolitan area, for example, it is not likely that every consumer has the option to choose among the alternatives. The providers may serve only limited areas of the city, and in any given neighborhood, there may be only one provider from which to choose. Furthermore, consumers may not be able to select more than one provider at the same time. If favored content is bundled with network access, and if the price for the bundle is such that consumers can afford a subscription with only one provider, then the result will be one set of affiliated content providers per household.[4]

The danger of restricting consumers' choice among information outlets is that, as individuals increasingly use the Internet to gather information to make important economic and political choices, they will have access to only one conglomerate's choice of content sources. In the transition to increased broadband connectivity and network convergence, restrictions on consumer choice will be more menacing.

---

[4] Yannis Bakos and Erik Brynjolfsson[1999] describe the economies of aggregation that can be expected in the distribution of information goods on the Internet.

Even where service providers allow connections (albeit lower of quality) with non-affiliated information providers, subtle differences in the quality of service—a few milliseconds delay here and a few milliseconds there—may be sufficient to condition users to steer clear of "slower" unaffiliated information providers.[5] On a larger time scale, researchers have shown that web content that downloads more rapidly is perceived to be "more interesting" than slower content.[6] Thus, it may not be necessary to do something so obvious as completely discourage users from visiting unaffiliated sites. But if the mechanisms for steering users to selected content providers are not so blatant as blocking, how can we determine if there is cause for concern?

This paper outlines a proposal for monitoring the potential effects of proprietary information spaces. We recommend that policy makers begin monitoring the QoS that broadband consumers are receiving. Then regulators will be able to determine if and how the user experience differs depending on whether users are viewing content from an information provider affiliated with the access carrier.

The broadband quality of service monitoring program we propose would consist of a small program running on consumers' computers and a server that collects and processes data from participants in the monitoring program. If the number of participants is sufficiently high, the analysis could yield data on the characteristics of one consumer compared to other customers of the same access provider, as well as comparisons of the service being offered by a different broadband access providers. Thus the program could provide information for several purposes. First, monitoring data would allow policy makers to determine if proprietary information spaces appear to be taking shape. Second, the data would provide consumers with an independent measure of the quality of service of competing broadband providers. By providing consumers with data about options available to them, the monitoring program would reduce the need of regulators to meddle in the affairs of broadband access providers through over-particular regulations of service characteristics.

---

[5] Jacko et al., [2000] for a study on the effect of delays on the perceived usability of on line resources.

[6] Ramsay, et al [1998]. In their experiments, Ramsay, Barbesi and Preece tested download times between two seconds and two minutes. They found that ratings for pages with downloads over 41 seconds were substantially lower than pages that downloaded more quickly.

As of this writing, any difference in performance between affiliated and unaffiliated content providers is probably accidental. The locations at which information providers have connected to the Internet are largely the legacy of choices made years ago with respect to location and service provider.

## 2. Quality of service measuring efforts in other industries

Quality of service monitoring as a complement or alternative to prescriptive regulation is becoming a favored option in a variety of industries. The advantages of quality of service monitoring over detailed rules are obvious. First is flexibility. Monitoring programs obviate the need for codification of soon-obsolete standards in the law books. Flexible regulation is particularly important in markets where consumer expectations evolve rapidly. Next, QoS monitoring enhances the ability of market forces to drive producers and consumers towards more efficient outcomes. QoS monitoring according to industry-wide standards allows consumers to better choose among a range of complex options. Monitoring programs provide comparison information to consumers who, in most cases, do not have access to as much information as providers. By requiring providers to make public quality of service information, regulators address the information disparity that exists between producers and consumers of complex, high-tech products. QoS monitoring leaves the decisions about what consumers want to the consumers, and it leaves decisions about the best way to satisfy consumers to the companies competing for their business.

Monitoring programs range from simple collection of complaints to specifying measurements and reporting standards.  A number of recent legislative efforts across several industries rely on the dissemination of information about service offerings to avoid detailed regulations.

Representative Anthony Weiner (D-NY) is sponsoring the "Cell Phone Disclosure Act of 2001," H.R. 1531.   Inspired by constituent complaints about the "dead spots," dropped and blocked calls in their cellular telephone service, the bill will require cell phone providers to publish information about the quality and coverage of their cell phone service.  The bill will direct the Federal Communications Commission to establish and administer a system, including a toll-free telephone number, for registering complaints on the performance of commercial mobile telephone services. The FCC is also to require each service provider to include in each subscriber's bill a statement informing the subscriber of such system.  Representative Weiner's bill proposes little that will allow consumers to assess or compare the services provided by competing wireless providers, however.  There are no uniform measurements—only the tallying of complaints.

There are several bills in the 107th Congress on airline passenger rights.  S. 483, the "Fair Treatment of Airline Passengers Act," requires airlines to disclose, "without being requested," the on-time performance and cancellation rate for any "consistently-delayed or canceled flight" at the time of reservation or purchase of a ticket.  The Act specifies that "consistently delayed" means that the flight has arrived late at least 40% of the time in the last three months; a "consistently-cancelled" flight is one that has been cancelled at least 20% of the time in the same period.

Similarly, the companion bill H.R. 1734, the "Airline Passenger Bill of Rights Act," requires airlines to disclose, "without being requested," the on-time performance and cancellation rate of a "chronically delayed" or "chronically cancelled" flight. In both cases, the new requirement that airlines volunteer the information will make for better-informed consumers. Now, critical information that consumers had to ask or search for will be given to them in the process of selecting a flight.

The attempt to redress consumer-vendor information disparities is reaching the health care industry as well. H.R. 2497, the "Managed Care Bill of Rights for Consumers Act of 2001," requires health plans to disclose the results of external reviews as well as annual data on plan compliance.

The legislation cited here represents just a few of the efforts in Congress to make it easier for consumers to make better-informed decisions. The emphasis on making information more available, rather than on adding new regulatory requirements, leaves producers and customers to decide the best mix of product attributes. This is the philosophy we recommend for the broadband access market. Rather than unworkable, detailed regulations, why not just ask providers to furnish information about the quality of the service they provide, and let the consumer select the best choice?

## 3. Broadband Architectures and Quality of Service

What aspects of network architectures affect a broadband access subscriber's perceptions of quality of service? The answer depends on the type of network involved. Network architectures for broadband access via DSL and cable modem differ slightly. But both types of networks include the following functional elements: (1) inside-the-house wiring; (2) a modem; (3) the "last-mile" access network between the modem and the service provider's local offices; (4) a modem in the service provider's network; (5) transport infrastructure across the service provider's network; and, at some point, (6) access onto what can be considered the public Internet.

The QoS perceived by a broadband consumer is in part determined by the speed with which the server at the remote end of the connection responds to service requests.  All told, the performance of a broadband access connection is a function of many network design and operational factors.  It is difficult for even the technically proficient consumer to make sense of the particulars of broadband quality of service.

The design and provisioning choices made by the access network operator will influence the quality of service enjoyed by subscribers to the system.  The following sections describe in detail those portions of the access provider's network that are most likely to affect the QoS as judged by the consumer.

**Last-mile access network**

Both DSL and cable modem services require a 'last-mile' connection between the subscriber's location and the telephone company's wiring center or the cable company's local head-end.  The wiring center (or central office) is the terminating location for the twisted pair copper line over which DSL services are delivered.  For cable systems, the "head end" is the location from which the cable operator distributes video and data signals over the (typically) hybrid fiber-coax network in local neighborhoods.  The table below illustrates the functional equivalents in the two systems.

| Common features of access networks | Cable data service | Digital subscriber line |
|---|---|---|
| Consumer premises equipment | cable modem | DSL modem |
| Last mile transport | fiber-coax or coax | Twisted pair copper line |
| Other end of the access line | Head end | wiring center/ central office |
| Provider premises equipment | Cable modem termination system (CMTS) | DSL Access Multiplexer (DSLAM) |
| Provider network | Various transport technologies; various servers | Various transport technologies; various servers |
| ISP or backbone | ISP's network, backbone connection | ISP's network, backbone connection |

Besides differences in the bandwidth of DSL and hybrid fiber coax networks, the major difference from the perspective of the user is the shared nature of the cable modem line and the dedicated nature of a DSL line. In a cable data network, many users—from hundreds to a few thousand subscribers may share the same data channels. On a DSL network, all of the capacity of the last mile is available to one subscriber only. But a comparison of the two broadband access technologies is not fair. Once the DSL user's data reaches the central office, it is dropped onto a network whose resources are shared among hundreds or thousands of DSL customers. (Charter Communications recently sued SBC for claiming in its advertisements for DSL service that only cable service slows down during peak usage periods, not DSL service.[7])

---

[7] "Charter sues SBC unit over ads" August 28, 2001. http://news.cnet.com/news/0-1004-200-6996741.html

Despite both DSL and cable data systems using shared resources, the problem of over-subscription to the last mile of cable networks is potentially a significant source of access quality degradation. At peak hours, a few users can saturate a network with enough traffic to limit the quality perceived by other users. The greater risk of neighbors causing network congestion in cable networks rather than on DSL lines is simply a matter of the statistical aggregation possible at points further upstream from the user. Of course, an under-provisioned provider network in a DSL (or cable) network can create the same problems that an oversubscribed cable network can create.

**CMTS or DSLAM equipment and service provider network**

Neither the CMTS nor the DSLAM itself is likely to be the source of sustained service problems. But network congestion on the provider's side of the CMTS or DSLAM can potentially cause quality of service problems for subscribers, as just described.

The servers operated by a service provider are another locus of QoS problems for broadband subscribers. The incoming and outgoing email servers, web proxy servers, caching servers, and news (NNTP) servers all directly affect the customer's perception of the quality of service that it is receiving from the service provider.[8] Should any of these servers suffer from congestion, the end user will perceive the quality of service to be dropping. For most subscribers, it will be difficult to distinguish between delays caused by congested local servers and congested local networks.

---

[8] In the following, we will use NNTP to indicate Usenet servers rather than the more colloquial "new server." Our intention is to avoid confusing an NNTP 'news' server with a web server that provides 'news' information.

**The autonomous system and quality of service**

In networking terminology, a network under the management of a single administrative entity is an *autonomous system* (AS.) The quality of service for any traffic that remains within an autonomous system is, therefore, entirely the responsibility of the administrators of that network. The quality of service of communications that cross boundaries between autonomous systems, on the other hand, are the responsibility of all of the AS administrators whose networks the path crosses.

In judging the quality of service that a broadband service subscriber is obtaining, therefore, we will have a much easier time drawing conclusions about those services that are accomplished inside the provider's AS. Mail and news services, in particular, tend to be provided by servers located inside the provider's network. The performance of a web session, however, may involve an arbitrary collection of service providers besides the subscriber's access provider.

In the following, we will be able to attribute any problems with quality of service for mail and NNTP services exclusively to the access provider. For the speed of web services, we will need additional information before we decide whether web surfing problems are being caused inside the access provider's network.

**Where should we make measurements?**

The broadband QoS monitoring program that we propose is implemented by measurements at the edge of the network—at the consumer's computer. As an alternative, we could propose more detailed measurements in the core of the network. But given our goals of monitoring the performance from the end user's perspective, there are several advantages to measuring the performance at the edge of the network.

A network operator has access to information on the status of equipment and links in its own network. With such information, the provider has a pretty clear idea of what is going on at a macro level. What the provider does not immediately have, however, is a picture of what end users are experiencing. So access to fine-grained information about network status, while valuable for managing the network, does not directly yield information on the consumer's satisfaction with the state of the network.

Another reason why we favor measurements at the edge is because a customer's ISP is typically not the only party involved in determining the user's QoS. If a subscriber is trying to use a path that crosses multiple autonomous domains, we would require all of the service providers in a path to submit network status information. In addition to the difficulty in coordinating the sharing of information, most providers would consider such information proprietary and would not divulge it to end users. In our monitoring proposal, we are concerned with the performance as perceived by end users, so information about the status of routers and servers in the network is not directly useful for us.

Finally, the main reason why we feel measurements at the edge are appropriate for our purposes is because we do not trust the service provider to report to the consumer on the quality of the service the consumer is receiving. We require an independent measure of the QoS, and thus a measurement taken at the consumer's premises.

**What measurements can we make?**

Given that we limit ourselves to measurements that can be made at the consumer's premises, what kinds of things can we measure to gauge the performance of the network connection that the consumer is using?

There are two general classes of measurements: low-level network phenomena and higher, application-level phenomena. To monitor the QoS of broadband access service, both types of measurements are valuable. The lower-level measurements give some hint about the contributing factors to QoS problems and may provide some diagnostic information for network operators; the higher, phenomenological application level phenomena tell us what the user experience is like. For the end user, application-level performance is paramount. Users care more about the overall time for an application to respond to a request than they care about the delays of individual packets. Consumers will be interested in low-level measurements to the extent that, with interpretation, they provide hints as to who is responsible for application-level problems with QoS.[9]

**Low-level measurements**

At the low end of the scale is the *ping* program. Ping is named after the sonar method of determining the distance to another ship. The program measures the time it takes for a pair of test packets to traverse the round trip between two hosts. Besides its ability to determine if a remote host is reachable, ping also provides a measure of the end-to-end delays on a path on the Internet. Data from pings must be interpreted carefully, however, since raw ping times do not represent the true round trip times between two hosts. First, the round-trip time as measured by ping includes processing time at the end points of the connection. If one of the hosts is experiencing heavy computational loads, the ping time will overestimate the amount of time it takes the network to carry traffic between the two hosts. Secondly, the paths that packets take on the forward and return paths are not necessarily the same. The ping time represents the combined travel delay, across one or two different paths, plus some processing time at the remote and local ends of the path.

---

[9] See David D. Clark's argument [1997, 218] that "the criterion that the user has for evaluating network performance is the total elapsed time to transfer the typical data unit of the current application, rather than the delay for each packet."

A second low level network test program is *traceroute*. Traceroute provides information on the path taken from the user's computer to a remote host. Traceroute also provides information about the contribution to total round-trip delays made by the network elements on the path between the user's computer and the remote host.

Just as with very low-level network performance statistics, ping and traceroute data do not directly indicate what the end user experience will be like. An underpowered web server, for example, might be able to respond to a ping in a timely fashion, but it may take a long time to generate the results from a search. The ping data from such a server would look good, though a person trying to view the web site might be frustrated by the delays in generating and rendering a web page.

**Application-level testing**

Compared to the low-level network data yielded by the ping and traceroute programs, application-level performance data measures directly describe the performance as perceived by the end user. . Application-level performance represents what is referred to in the user research literature as *system response time*, or the time that elapses between a request for an action and the time that the computer completes the action. An example of an application-level measurement is the elapsed time to download a web page and all of the graphics on it, from the time of the request until the page is fully downloaded. Other examples of application-level measurements are the amount of time required to download or upload email messages to/from an email server, or the time to download message lists or messages from a news server.

Application-level phenomena represent the cumulative, perceptible effect of the various packet-level phenomena. Where the components of the application-level phenomena are under the control of the user's ISP, then we have no difficulty in assessing responsibility for any problems with QoS. Where some of the delays are the result of conditions off of the ISP's network, however, we must be careful in attributing blame for any extraordinary delays. With the help of ping and traceroute data, however, we can begin to assess whether unusual delays are being caused by problems inside or outside of the ISP's network.

**A note on TCP and measuring data rates**

We distinguish between the effective rate at which a file can be transferred across a network and the raw data rate of the channel. The differences between the two numbers are important for the user. The raw data rate of a channel merely indicates the rate at which the modems in the channel send and receive bits. The effective rate is lower than the channel rate for a variety of reasons. First, there are overhead bits in the transmission channels we are discussing. We are interested in the net data rate of an IP channel. Hence we do not count any IP and TCP overhead data in the effective data rate of the channel. Second, a channel may be shared with other users, as is a cable data network. So while a cable modem might send its data at 30-40 Megabits per second, it is unlikely that any individual user would ever enjoy that speed on a cable data network.[10]

---

[10] First, the network is shared, so no single user is likely to be able to use the entire bandwidth for any extended period of time. Second, the cable company's CMTS might never allocate that much bandwidth to a single user, even if there is capacity available to do so. Third, the user's computer might not have the ability to send data at that rate (if, for example, a computer has only a 10 Mbps Ethernet card, there is no way that it could sustain transmissions at the cable data network's bit rate. Other components in a user's computer, such as the hard disk, might also make it impossible to reach such speeds.)

Whatever the maximum theoretical data rate of a channel, there are characteristics of the Transmission Control Protocol that we must consider when estimating the maximum data rate available to a broadband subscriber. To avoid the risk of fast hosts swamping slower hosts and routers with too much data, TCP uses a scheme known as "slow start" to ramp up the rate of transmission. If a computer sending data via TCP fails to receive acknowledgements, it assumes there is congestion on the network and cuts back on the rate at which it is sending data. The slow start algorithm slowly increases the rate at which the computer sends data until some link in the path, often (but not always) the last mile link at the user's end—reaches saturation and can handle no more data.

To measure the maximum practical, net data rate available to a user at a given time, then, it is best measure the time to download as large a file as is practical. That way, under normal conditions, the slow start algorithm will have an opportunity to stabilize on the maximum practical transmission rate that the network can handle. Also, by testing with as large a file as possible, we minimize the effect of the slow start on the estimated data rate (as calculated by number of bits/number of seconds to completion.)

The effect of the slow start algorithm on the speed at which files can be downloaded off of the Internet for one user can be seen in Figure 1. For this broadband subscriber, the slow start algorithm tends to reach a rate limited by the speed of the DSL service for files over the size of 250-300 kilobytes. For smaller files, the slow start algorithm does not max out the speed on the DSL service. For larger files, the average rate for a file transfer asymptotically approaches the speed of the DSL circuit, minus TCP/IP overhead and any layer 2 overhead (such as might be necessary for PPPoE.) In this case the subscriber has a 640 kbps raw data rate DSL line, and the throughput looks like it would never exceed about 500 kilobits per second.

In the chart, each marker indicates a file; for each file, you can determine the average rate at which it downloaded off of the net.

**Figure 1. TCP slow start.  Average throughput per file, by file size**

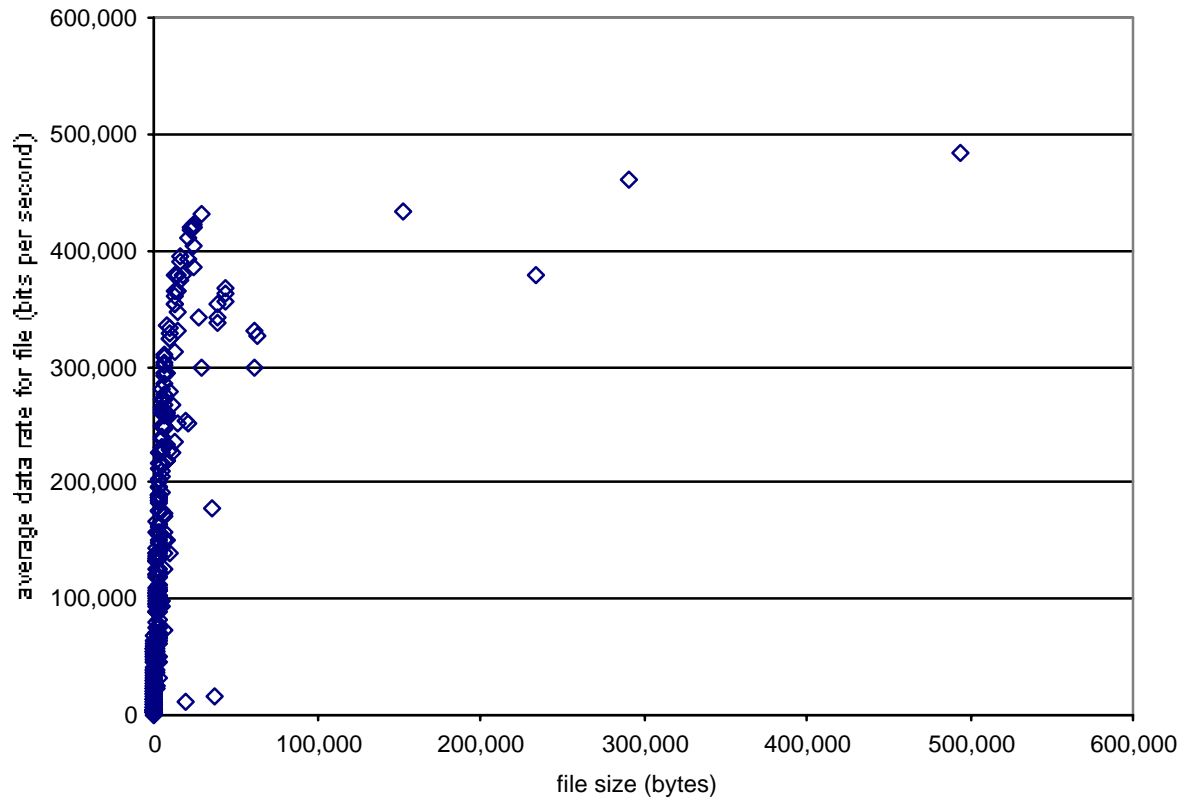

**Dimensions of Quality of Service**

What are the components of quality of service for a broadband connection?  The most obvious, and the one highlighted in marketing information, is the speed of the connection. But the user's view of "speed" of an Internet connection is a function of the *data rate* and *latency* of a connection. Each of these dimensions can be independently assessed using the appropriate tool.  The higher the data rate, and the lower the latency, the more responsive network applications will appear to the user.

The data rate is simply the frequency with which bits travel along a line. Latency is the end-to-end delay on a path on the Internet. Both signal propagation delay and processing times contribute to network latency. At the speed of light in fiber optic cable, a cross-country path should take at least 56 ms. In practice, however, round trip delays across country can be expected to start near 100 ms. Hence processing delays constitute a significant, if not predominant, share of latency. Computational delays comprise the delays introduced by routers and other network elements as well as the computational delays on the hosts at the ends of the connection. Internet applications with lower latency appeal to the user.

A networking layer statistic that affects user perceptions of the quality of a service is the packet loss rate. The Internet protocols provide a "best-effort" communications medium, meaning that there is no guarantee than any individual packet will make it safely to its destination. Higher level protocols, like TCP, or applications, can provide a reliable connection using the best-effort IP networking protocols. If a packet is lost, for example, TCP will insure that it is sent again. Packet losses add to the end-to-end delays perceived by the user.

The ping program has the ability to measure the probability of packet loss along a path. By repeatedly pinging a remote site and comparing the number of successful to unsuccessful ping packets, a user's computer can detect when there might be problems with packet loss. (The ping program can overestimate the probability of packet loss, since ping packets are more likely to be dropped than normal IP packets. But if there is a serious packet loss problem, the ping program will detect the problem.)

**An architecture for a broadband QoS monitoring system**

Now that we have outlined the measurements that can be made of broadband access, we can outline the architecture and financial underpinnings of a broadband quality of service monitoring program. The components of the system are a large set of subscribers running the software on their computers and a central server that collects and processes data collected by the participants' computers.

Because the data collected by any individual subscriber at any given time is contingent on a number of unknown factors, it would not be wise to rely on any single set of measurements to characterize the quality of service available to any individual broadband subscriber. There is "noise" in the data—random variations that could mislead analysts attempting to make inferences about the quality of service. But multiple observations, made over an extended period of time, would allow analysts to filter out the non-systematic noise in the data. Comparisons of access times for groups of customers of different access providers would in time show if a systematic bias in the quality of access to selected content sources was taking shape.

Who would pay for these monitoring efforts? Experience shows that consumers are often unwilling to pay for the verification of the services that they received. Unless consumers are planning on making a major purchase like a car, for example, they are not willing to pay for services such as those offered by *Consumer Reports.*

Service providers might be willing to fund broadband quality of service data collection and analysis, however. The incentives for providers to pay for the services are clearer: if a provider feels that it is within the range of promised service quality, monitoring can be used to prove that customers were receiving the quality of service promised under a service level agreement (SLA). If a provider fails to meet its obligations under an SLA, then it might like to have data detailing the degree to which it undershot its goals, so that it might verify to its customers that it is compensating its customers to the degree promised in the SLA.

Regardless of whether consumers pay directly for broadband monitoring, they will find a service provider's offerings more attractive if the provider participates in an independently managed monitoring program. If consumers have a choice of buying an unverified service vs. a verifiable one, they are, other things being equal, more likely to select the verifiable option. Service providers who offer a quality product will thus find that a monitoring program will reinforce any claims it makes in advertising.

Most significantly, however, service providers may want to sponsor broadband quality of service monitoring so that they can provide regulators with data that demonstrate compliance with rules. And if service providers do not build walled gardens around favored content, monitoring data can also show that there is no need for regulatory intervention in information markets.

### Monitoring and Service Level Agreements (SLAs)

A Service Level Agreements (SLA) is a contract between a service provider and a customer in which the provider specifies the characteristics of the service that it will provide to customers. SLAs may specify the "up time" or availability of the network, the data rate, the latency, or other quality of service properties. The SLA spells out the measures that the provider will take in the event that it is unable to deliver the service specified in the SLA. The provider may guarantee that it will restore service within some number of hours, or it may specify the restitution that it will make to subscribers if it is unable to meet the quality of service goals.

While customers would like SLAs to cover every contingency, providers can make promises only for those service characteristics that they control. So SLA promises concerning data rate, latency and availability apply only to traffic that remains on the service provider's network. For example, WorldCom's SLAs for its commercial customers provides for the refund of one day's fees for every hour of network outage. WorldCom also promises latencies of 65 ms or less between "WorldCom-designated inter-regional transit backbone routers" in the contiguous United States. WorldCom guarantees delivery of at least 99% of packets "between UUNET-designated Hub Routers in North America." Finally, for customers with connections to the WorldCom network in the United States, the SLA specifies that WorldCom promises that it will notify customers if it detects that the customers network is not accessible from the WorldCom network.[11]

Currently, SLAs such as the WorldCom agreement described above are available for large customers only. Consumers are not yet able to get service providers to promise anything nearly as comprehensive for their broadband access. In fact, some broadband providers seem to go out of their way not to raise customer expectations. In its DSL sales literature, Verizon states:

> Due to the sophisticated nature of DSL, Verizon Online cannot guarantee uninterrupted or error-free service, or the speed of your service.[12]

With evasive statements about service quality like this, how is the consumer to choose a broadband service provider? If consumers had access to monitoring data on Verizon's DSL service, they would not have to try and make sense of Verizon's non-committal language. They could just go to the data and figure out if Verizon offered the best broadband value.

---

[11] As of 9/24/01, the SLA is described at
http://www.worldcom.com/terms/service_level_guarantee/t_sla_terms.phtml

[12] As of 9/24/01, "Verizon Online DSL: About Speed"
http://www.bellatlantic.com/infospeed/more_info/about_speed.html

## 4. Broadband QoS monitoring technology

To demonstrate the feasibility of measuring broadband quality of service, we constructed a program that runs on Microsoft Windows machines and periodically measures the quality of service at both the application and network phenomena levels. This program does not include all of the features that we would envision for a commercial-quality monitoring program. Nor have we developed the software that would run on a data-collection and analysis server discussed in section 2. Still, we find that a simple program demonstrates the feasibility of a monitoring program to detect differences in the quality of service based on access provider's network connections and affiliations.

The application can be configured to measure email performance, web performance, or news (Usenet) performance at selected intervals, or the program can be made to conduct tests manually.

There are millions of different types and configurations of computers on the Internet. There is no reason to believe that any two computers would be able to render a web page in the same time as each other. To avoid getting tangled up in measurements of the capabilities of one computer over another, the software measures only the time to download the data onto the user's computer. In practice, the system response time would include the time required for the computer to render the results of a request on the screen. Hence, the measurements that we present here slightly underestimate the delays that an end user would perceive.

Another shortcut the program takes is that it downloads the data but does not store it to disk. This allows us to avoid adding unknown amounts of disk access time to the download times. Thus, within the limitations of the Windows operating system, the download times measured by the monitoring program are as close as possible to what can be considered the best estimate of the true time to download the files.[13]

### Email tests

The program conducts tests of both incoming (Post Office Protocol, or POP) servers and outgoing (Simple Mail Transfer Protocol, or SMTP) servers. The metric used to measure the performance of the email service is the amount of time required to transfer messages to and from the mail servers. For incoming mail, the monitor measures the total elapsed time from the request to download an email message until the message is completely received. For outgoing mail, the monitor measures the amount of time required to upload a message to the SMTP server.

Because of TCP slow start, it is likely that the speeds measured for very small email messages will not represent the speed at which the server and network are capable of delivering mail. For that reason, the monitoring program has the ability to send itself a large (>300 kilobyte) message. For broadband services, this file size allows the TCP link between the user's computer and the mail servers to approach their maximum.

Access providers run their own email servers for consumer and small business customers and bundled email service with access service. The email servers are typically located on the access provider's network, so any problems with the quality of service can be entirely attributed to the service provider.

---

[13] Windows imposes its own uncertainties into the process, including inaccuracies in the system timers, the inability of the monitor to block other applications from taking over the system resources while it is conducting tests. The timing scheme is also susceptible to quirks in the Windows networking APIs.

**Web service tests**

The monitoring program also tests the rate at which it is possible to download HTML and other files on a web page. The monitor first times the download of the HTML file for a web page. Then it scans the HTML for embedded elements, then times the download of each of the elements. (As would a web browser, the monitor downloads each element only once, no matter how many times it appears on the page.) When all of the elements have been downloaded, the monitor calculates the total time to download the web page. Depending on how the preferences for the program are set, the monitor reports either all download times or the aggregate time for each page downloaded.

The monitor program determines what web pages to visit by downloading a list of URLs from a web site run by one of the authors. For the data represented below, the list directed the monitoring program to download the home pages of www.nytimes.com, www.cnn.com, www.usatoday.com, and a fourth URL selected in round-robin fashion from a list of URLs pointing to large files (between 200 kilobytes and 1 Megabyte.) As with the email case, the selection of a large file allowed the program to ramp up the TCP connection to its maximum.

There are two shortcomings in the implementation of the monitor program that collected the data discussed below. The first concerns the use of JavaScript on web pages. Some sites use JavaScript to dynamically generate the layout and links for a page. Our simple proof-of-concept program does not have the ability to parse and perform JavaScript functions, so we were forced to avoid sites that relied heavily on JavaScript.

The second shortcoming is yet another consequence of the slow start algorithm. A complex web page is likely to have a large number of small graphical elements on it. The monitor downloads these files seriatim, while a real web browser would download these files in parallel, using separate threads of execution. By downloading files in parallel, the browser can more efficiently use the link to the Internet when the individual streams are limited in speed by the slow start algorithm. When a browser is connected to a server that does not support 'keep-alive' connections, each page element requires a new TCP connection to start from scratch.

Whether the presence of web caching invalidates the download time measurements depends on whether we are interested in measuring the time to download the files from the server or whether we are interested in measuring the user's experience of the download. Our interests are clearly in the user experience, so the question of whether content is cached locally or not is not important. It is entirely possible that an access provider could provide more favorable caching for selected content, pushing it further out to the subscribers, while less favored content is cached, if at all, at a considerable distance from the subscriber.

### Usenet server tests

Finally, the monitor is able to measure the performance of NNTP Usenet servers. When conducting tests of the subscriber's NNTP server, the monitoring application downloaded a list of newsgroups from a website run by one of the authors. The newsgroups selected for testing were binaries groups, since they are likely to carry large messages.

When conducting NNTP tests, the monitoring program first downloaded a complete list of groups carried by the NNTP server. Premium NNTP servers that cater to heavy users can carry more than 60,000 groups; less ambitious news servers still carry thousands to a few tens of thousands of groups. The list is not just a dump of a text file—it requires a lookup in the database of the number of articles available in each group. Downloading the full group list, therefore, takes some time and requires thousands of exchanges between the server and the user's computer.

Next, for each of the newsgroups in the list obtained from the author's web site, the monitoring program downloads up to 500 article headers. This process also is more complicated than a simple file transfer; for fast access lines the speed of the group listing is likely to be less than the maximum line speed.

Finally, after downloading the list of up to 500 articles from the selected group, the monitor searches through the list for the N largest messages, where the number N is obtained from the same download as the list of selected newsgroups. Large postings in binary newsgroups can run into the Megabytes, so the download speed can approach the maximum speed possible on the broadband subscriber's line.[14]

For the purposes of this report, we focus on the web performance tests, since they best exemplify the feasibility of testing accessibility of information across network domains.

## 5.  Preliminary results

We have collected data from the authors' computers as well as from computers of a number of our colleagues. From a preliminary analysis of the data, we can show how a broadband access quality of service monitoring program would be able to detect the early signs of proprietary information spaces. We find no cause for concern about proprietary information spaces in our data. We merely point out differences in the performance observed in accessing selected web sites from selected broadband service providers.

---

[14] Some ISPs cap the speed at which subscribers can download data from NNTP servers. See "Southwestern Bell DSL customers sue over bandwidth," August 17, 2000, http://news.cnet.com/news/0-1004-200-2549826.html for reports of caps on SBC's DSL service. If there is an artificial cap on the news server, then the speed reached while downloading articles from the server would not max out near the broadband access line's capacity.

As an example of the type of tests that might show signs of proprietary information spaces, we configured the monitoring program to test the download times for the home pages of three major US news sources: the New York Times, CNN, and USA Today. The monitor measured the download times of the HTML files and all graphic elements on the page. For the scatter diagram in Figure 2, the x-axis represents the size of the files downloaded; the y-axis represents the data average rate at which each file arrived (total size in bits divided by elapsed seconds.) The asymptotic shape of the curve depicts the effect of the slow start algorithm, discussed above.

**Figure 2. Performance by content source (640 kbps DSL line)**

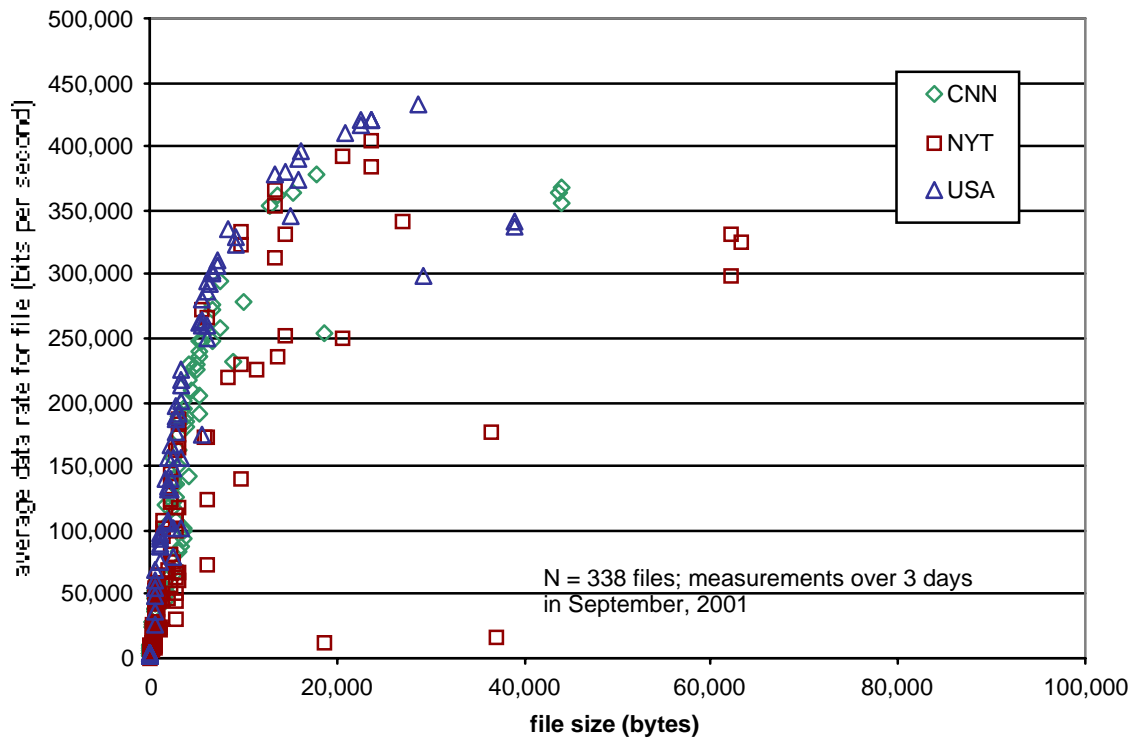

Data for the three content sources are plotted separately. A few patterns emerge from the data.  For this broadband subscriber, the files from the USA Today site appear to populate the frontier limit of the slow start algorithm.  Files from CNN are close behind in average data rate (remember to judge distance from the USA Today results by looking at the vertical distance between two points with the similar x-values.)  Finally, the data from the New York Times appears to be scattered all over the chart.  We cannot be sure if the variance in the www.nytimes.com data is due to problems on the Internet or problems with the nytimes server.  By comparing these results with those from other participants in the study, we can begin to make inferences about the source of the problem.

Data from a second DSL subscriber (with a greater than 1 Mbps line speed) show an unexpected distribution of file size and data rate.  (See Figure 3.) The slow start ramp up appears to stall at about 125 kilobits per second for files up to about 25 kilobytes in size.  But large files download at rates that approach the speed of the DSL line.

**Figure 3. Data rate by content source for a second DSL subscriber**

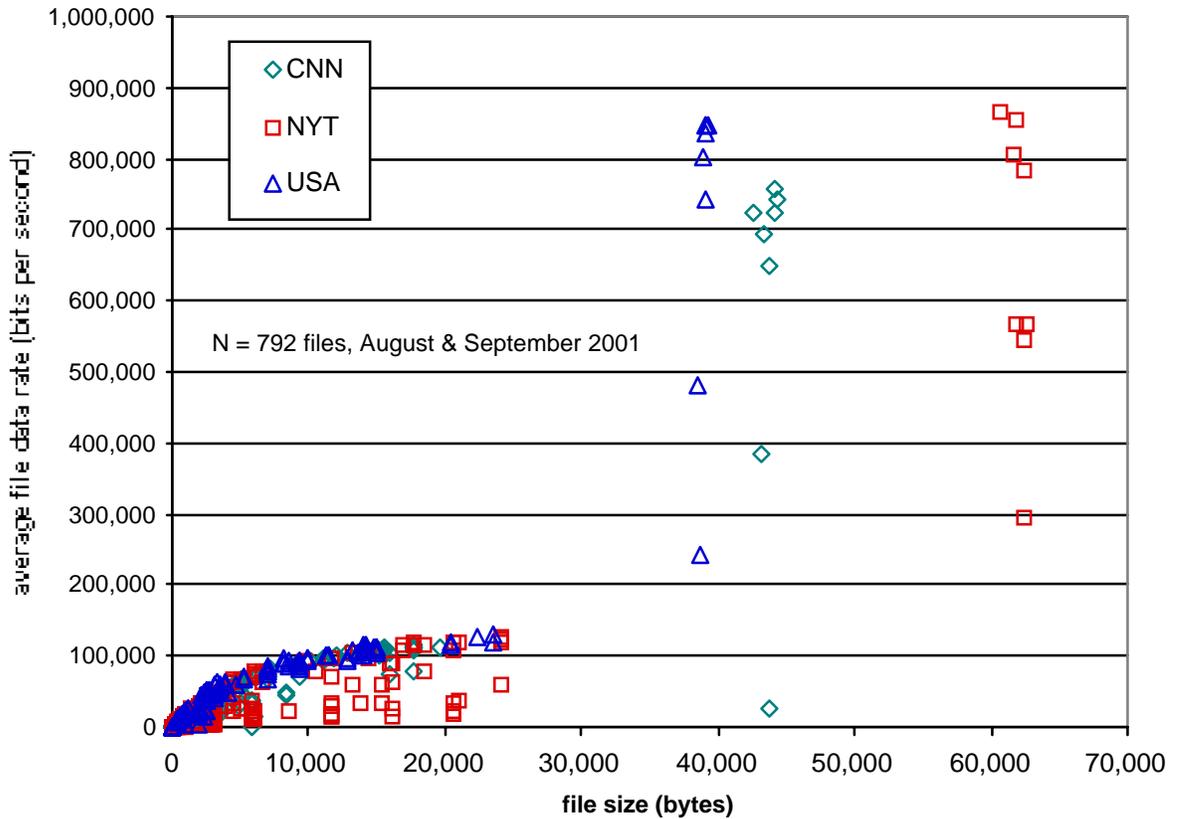

This shape is the opposite of what might be expected from caching—the larger HTML files, some of which contain a "no-cache" directive, are arriving at rates that might be expected, but the smaller files are arriving at a rate substantially below what might be expected.

Finally, the last set of data represent the performance measured by a cable modem subscriber. See Figure 4. In the last data set, there is a much larger spread of data from all three content providers, though the USA Today content appears to be arriving more consistently near the maximum speed than content from other providers.

**Figure 4. Performance by Content Source, Cable Modem subscriber**

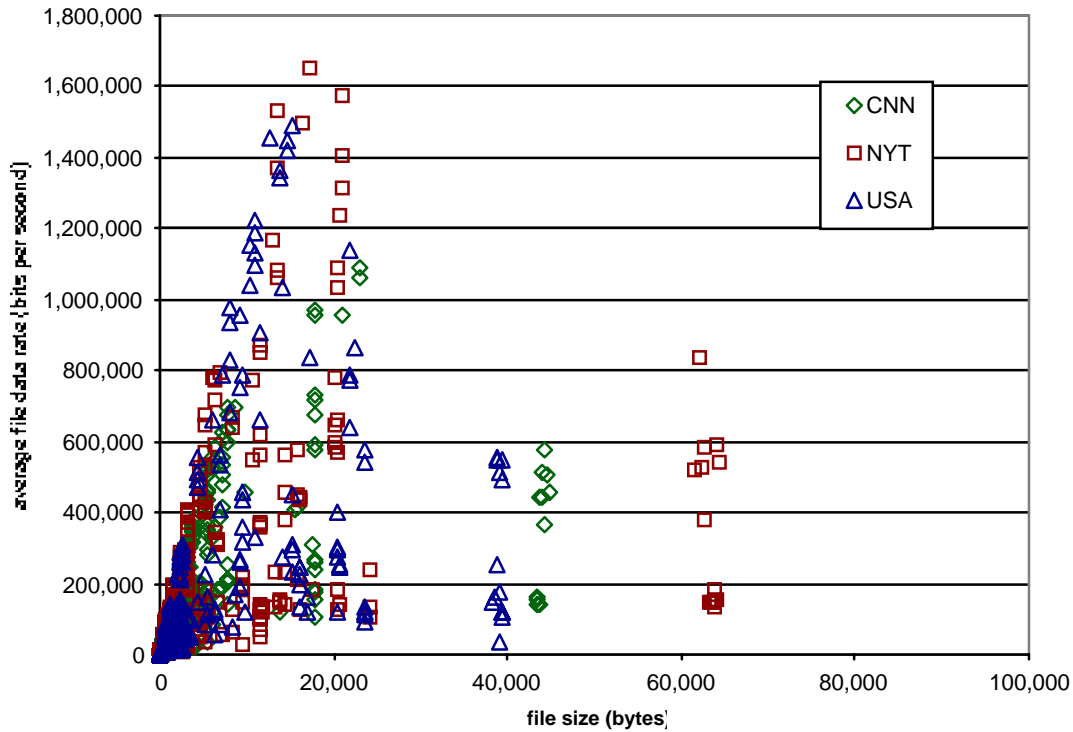

For the cable modem subscriber, the maximum data rates are substantially higher than for the DSL subscribers, though the data rates for the larger HTML files are comparable with those experienced by the DSL subscriber with the 1 Mbps service.

Note that the data for the nytimes.com files are scattered in all three cases, through the data for the other sources do not vary as much (for a given file size) for the other two broadband subscribers. It appears that the variability in the nytimes.com files must come from either the nytimes.com server or caching servers, or congestion on the network near the nytimes server.

The differences that we find in terms of speed of access for the sites in question may be a distinction that doesn't make a difference. That is, the differences we observe with our objective timing instruments may be too small to make a difference for the subjective instruments of the human brain. We encourage our colleagues in the industry to consider sponsoring a series of studies on the relationship between variations along various dimensions in broadband quality of service space and subjective and behavioral responses to those differences. We remind our colleagues to keep in mind that tests of subjective ramifications of objective differences in quality of service need not be consciously observed by users to make a difference in behavior. If small variations in the availability of content sources can affect the users' emotional or frustration levels, then the small objective differences may condition users to prefer (imperceptibly) faster affiliated content sites.

## 6. Conclusions

> "One of our problems is that our rules are not drawn to deal with competitive market situations. Consumer protections are kind of thin."
>
> California PUC Commissioner Carl Wood on complaints about Pac Bell DSL service.[15]

The measurement program we suggest is a first step toward metrics for broadband access QoS monitoring. As California PUC Commissioner Wood notes, regulators are ill-equipped to look after consumer interests in competitive markets. A quality of service monitoring program would provide consumers with the information necessary to make informed choices among industry offerings. Regulators could use the same data to judge the state of competition (and hence the appropriateness of regulation) in broadband access markets. And service providers could use the techniques to benchmark and manage service quality on their networks. We welcome conversation concerning monitoring techniques and metrics and hope that by better informing consumers, we will promote the growth and performance of broadband access services.

---

[15] San Francisco Examiner, Sunday June 11, 2000. http://www.sfgate.com/cgi-bin/article.cgi?f=/examiner/archive/2000/06/11/BUSINESS5860.dtl

The FCC continues to take a hands off, don't-fix-it-if-it-ain't-broke approach to broadband access markets. While such a policy has sufficed for the Internet to date, the deepening links between access and content providers concern us. Access providers without a source of content may find themselves unable to compete with vertically integrated access providers. The loss of competition in the access market would have predictable implications for the market. Consumers could find the range of choices in content restricted by the same vertically integrated access/content providers.

Under Chairman Kennard, the FCC adopted a policy of "vigilant restraint" regarding open access to cable data systems. Under Chairman Powell, the Commission is unlikely to take action until it sees that harm is being done by distortions in the marketplace. While we see the rationale of their position, we wonder if it is necessary to wait until after a small number of vertically integrated conglomerates put a stranglehold on the access markets. We counsel a middle ground—forbear from imposing regulations, but collect data in the meanwhile. If the data show that proprietary information spaces are taking shape, then the Commission can act before competing voices are driven from the market.

## Bibliography

Bakos, Yannis and Erik Brynjolfsson. 2000. "Bundling and competition on the Internet." *Marketing Science*, **19**(1). 63-82.

Clark, David D. 1997. "Internet cost allocation and pricing." In Lee W. McKnight and Joseph P. B ailey, eds., *Internet Economics*. Cambridge: MIT Press.

Jacko, Julie A., Andrew Sears and Michael S. Borella, 2000. "The effect of network delay and media on user perceptions of web resources," *Behaviour and Information Technology.* **19**(6), 427-439.

Stevens, W. Richard. 1994. TCP/IP Illustrated, volume 1. Reading: Addison Wesley.

Ramsay, J. A. Barbesi and J. Preece. 1998. "A psychological investigation of long retrieval times on the World Wide Web.", *Interacting With Computers*, **10**, 77-86.